\documentclass[12pt,a4]{article}
\usepackage{amsfonts}
\usepackage{amssymb}
\usepackage{amsmath}
\usepackage[T2A]{fontenc}
\usepackage[cp1251]{inputenc}
\usepackage{hyperref}
\usepackage{xcolor}

\usepackage{graphicx}

\numberwithin{equation}{section}
\begin{document}

\title{A way to constrain a graviton mass from astronomical observations}
\author{
Alexander~F.~Zakharov\\ 
 National Research Center ''Kurchatov Institute'',\\ Moscow 123182,  Russia;\\
Bogoliubov Laboratory for Theoretical Physics,\\
 JINR, Dubna 141980, Russia \\
Predrag Jovanovi\'c\\
Astronomical Observatory,
Volgina 7, \\
{P.O. Box 7}4, 11060 Belgrade, Serbia\\
{Du\v{s}ko} Borka and 
Vesna Borka Jovanovi\'c\\
{Department of Theoretical Physics }\\
{and Condensed Matter Physics (020),}\\ 
Vin\v{c}a Institute of Nuclear Sciences\\ {- National Institute of the Republic of Serbia,}\\ 
University of Belgrade, P.O. Box 522, 11001 Belgrade, Serbia \\
{\small Keywords: General Relativity, Alternative theories of gravity, Massive Gravity}, \\
{\small Modified Gravity, Graviton, Gravitational radiation, Black Holes}} 

\maketitle

\begin{abstract}
Along with a whole range of alternative theories of gravity, variants of the massive theory of gravity, i.e. the theory of gravity, in which the graviton has mass, have been actively discussed in recent years. Theorists have proposed versions of massive gravity theories that address the shortcomings of early versions of such theories.
Astrophysicists and experimental physicists have been discussing limitations on the graviton mass from various astronomical observations. In particular, in the first LIGO paper, where the discovery of gravitational waves from binary  black holes was reported, a limit on the mass of the graviton was obtained from the analysis of the profile of the gravitational wave signal. In the paper graviton mass constraints were obtained by analyzing the trajectory of a bright star in the vicinity of the center of our Galaxy, using observations from the GRAVITY and Keck groups. Briefly other astronomical ways to limit a graviton mass were discussed.

\end{abstract}
 
\section{Introduction}

The general relativity (GR) was established in November 1915 after intense discussions between A. Einstein and D. Hilbert {\cite{Vizgin_01,Logunov_04}} (a short overview of important aspects of the  cosmology development   is given in \cite{Zakharov_25_Baldin}).
For more than 110 years of its development, GR has successfully passed many observational tests and a huge number of predictions based on this theory have been confirmed.
Nevertheless, there are problems in clarifying the nature of dark matter and dark energy, and according to a number of theorists, these problems can be solved with the help of modified general relativity. In addition, despite numerous attempts, a quantum theory of gravity that is consistent with both the Standard Model and the Einstein's classical theory of gravity has yet to be developed. 

In one of his last papers F. Dyson noted that graviton detection is extremely hard issue \cite{Dyson_13}. However, if graviton is massive
there are different ways to limit its mass as it was discussed below. The first version of massive gravity theory was considered
in \cite{Fierz_39}. In 1970, it was found so-called the discontinuity \cite{vanDam_70,Zakharov_70, Iwasaki_70} or in other words, properties of the
massive gravity theory for $m_g \rightarrow 0$ and $m_g=0$ are different ($m_g$ is a graviton mass). However, due to the Vainshtein screening \cite{Vainshtein_72,Babichev_13}
this pathology may be not so dangerous. Other pathology was found in {the} 1970s, namely the Boulware -- Deser  (BD) ghosts in  non-linear Fierz -- Pauli theory were discovered
\cite{Boulware_72a,Boulware_72b}. For a long time, it was believed that the presence of ghosts in a massive theory of gravity was the main obstacle to create a self-consistent massive theory of gravity. {In spite of the found pathologies, the massive gravity theory was actively discussed in the last decades, see, for instance, a review \cite{Rubakov_08}.}

A class of ghost-free
massive gravity has been proposed in papers \cite{deRham_10,deRham_11} such a theory are called now the de Rham--Gabadadze--Tolley (dRGT) gravity model (see also more comprehensive reviews \cite{deRham_14,deRham_17}).
A contemporary review of a modern status of massive theory of gravity was presented as a chapter in a recent book 
\cite{deRham_24}.

It is believed that there are (supermassive) black holes in the centers of a large number of galaxies. Nevertheless, it is desirable to have additional evidence confirming the existence of black holes in galactic centers \cite{Genzel_24}.
To study the properties of the gravitational field in the Galactic Center, astronomers observed the motions of bright objects in its vicinity.  In particular, in recent decades, astronomers have achieved remarkable results by observing bright stars in the vicinity of the Galactic Center. 
One group led by A. Ghez (UCLA) uses Keck telescopes at Hawaii.
Another group led by R. Genzel, basically formed by researchers from European Sourthern Observatory (ESO)
and Max Planck Institute for Extraterrestrial Physics or shortly MPE, (both institutions have their headquarters in Garching near Munich) uses 8.2 m Very Large Telescopes at Paranal in Chile.
In November 2025 the advanced version of the interferometer GRAVITY+ consisting from four VLT telescopes started its observations\footnote{\url{https://www.mpe.mpg.de/8106898/news20251028}.}.
Previously, astronomers used the initial version of GRAVITY interferometer for about a decade and made a number of remarkable discoveries with it \cite{Genzel_21}, for instance
the GRAVITY collaboration found that gravitational redshifts for  S2 star orbit in May 2018 is in correspondence with GR expectations
\cite{GRAVITY_18,GRAVITY_19} (these conclusions have been confirmed by the Keck group \cite{Keck_19}).
Several years ago the GRAVITY collaboration reported a detection of the Schwarzschild precession for the S2 star orbit near GC \cite{GRAVITY_20}.
This result is extremely important, because despite the fact that the Schwarzschild precession of an elliptical orbit was firstly considered by A. Einstein to explain the anomaly of Mercury's orbit in 1915, this observational result of the group actually confirms the universality of the laws of gravity (using the example of the Schwarzschild precession) in various astronomical objects, such as the solar system and the vicinity of the Galactic Center.

It should be noted that even before the confirmation of relativistic effects, the parameters of the total mass distribution in the vicinity of the Galactic Center can be limited  from the analysis of the motion of bright stars \cite{Nucita_07,Zakharov_07}, {therefore, Galactic Center model can be clarified}.
In 2024 the GRAVITY collaboration improved constraints on extended mass distribution near the Galactic Center with bright star orbits
\cite{GRAVITY_24}.

{In addition to verifications of relativistic predictions, parameters of alternative theories of gravity can be constrained using results of observations of these bright stars}.
{For instance, some time ago}, theorists considered the theory $f(r)=R^n$ ($n=1$ for GR case) as a suitable alternative to the conventional GR since it shown it was possible to fit an accelerated expansion of the Universe and flat rotation curves in the framework of the theory. However,  considerations of celestial mechanics in Solar system
\cite{Zakharov_06} and  trajectories of bright stars near the Galactic Center \cite{Borka_12,Zakharov_14}
put strong constraints on parameters of this theory. 
In \cite{Zakharov_18b} it was shown that observations of bright stars could be useful to constrain a (tidal) charge of the supermassive black hole
at the Galactic Center. Some time ago, it was {proposed} to substitute supermassive black hole in the Galactic Center with  a dense dark matter ball
\cite{Ruffini_15,Becerra_21}. Really, test bodies move along elliptical trajectories inside such a ball, however,
properties of the trajectories are different from observed ones \cite{Zakharov_22,Zakharov_22b}. 

Taking into account the GRAVITY discovery of the Schwarzschild precession for S2 star orbit constraints on parameters of extended gravity theories were found \cite{Borka_21}.

\section{Ways to limit graviton mass from astronomical observations}

If graviton has a mass then as it was noted by C. Will \cite{Will_98,Will_06,Will_14}  its mass can be estimated from observations of gravitational waves
as it was done later by the LIGO -- Virgo -- KAGRA (LVK) collaboration\footnote{Two interferometers
LIGO  with 4~km arm lengths are located in Livingston (Louisiana state, USA) and Hanford (Washington state, USA), the interferometer Virgo with 3~km arm lengths  is located in Cascina (Italy), the underground Japanese interferometer KAGRA with 3~km arm lengths 
is located in the Kamioka mine.}. Even in the first publication of gravitational wave detection by LIGO interferometers, the LIGO -- Virgo collaboration declared not only the discovery of gravitational waves from binary black holes with masses around $30~M_\odot$\textbf{,} but they also constrained graviton mass as $m_g < 1.2 \times 10^{-22}$~eV$/{\rm c^2}$  \cite{LIGO_16}
(this estimate was consistent with a preliminary constraint  $m_g < 2.5 \times 10^{-22}$~eV$/{\rm c^2}$  done by C. Will in \cite{Will_98}).
Analyzing data collected in the second gravitational wave catalogue the LIGO -- Virgo collaboration tightened their graviton mass limit $m_g < 1.76 \times 10^{-23}$~eV$/{\rm c^2}$  \cite{LIGO_21}.
Using data from the third gravitational wave catalogue the LIGO -- Virgo collaboration updated their graviton mass limit  $m_g < 2.42 \times 10^{-23}$~eV$/{\rm c^2}$  \cite{LIGO_21b}.

To explain dark matter and dark {energy} phenomena in last decades theorists proposed a number of alternative theories of gravity and some of these theories have  Yukawa gravity in a weak gravity limit \cite{Capozziello_08,Capozziello_11}. In \cite{Borka_13} it was shown that parameters of gravity theory with a Yukawa
potential can be constrained using an analysis of trajectories of bright stars moving around the Galactic Center (these observational data were obtained by Keck and VLT collaborations).
Later, graviton mass constraints were found from an analysis of the trajectories of bright stars \cite{Zakharov_16,Zakharov_16_Quarks} and this constraint is $m_g < 2.9\times 10^{-21}~{\rm eV}/{\rm c}^2$
slightly worse than the estimate that was given in the first publication on the discovery of gravitational waves, given by the  LIGO -- Virgo collaboration.
Using new data the Keck group improved the estimate and these authors obtained  $m_g < 1.6\times 10^{-21}~{\rm eV}/{\rm c}^2$ \cite{Hees_17,Hees_17_Moriond}.
Our graviton mass constraint obtained from an analysis of S2 star trajectory \cite{Zakharov_16} together with other graviton mass bounds (including LIGO -- Virgo ones)  was listed
in PDG manual\footnote{\url{https://pdg.lbl.gov/2025/listings/rpp2025-list-graviton.pdf}.}.
 A rough estimate of perspectives to improve graviton mass bound with analysis of bright star orbits was given in paper \cite{Zakharov_18}.
Taking into account the Schwarzschild precession for S2 star orbit and using MCMC simulations for different orbits of the bright stars moving around the Galactic Center
 new constraints on Yukawa gravity parameters and on  graviton mass were found in \cite{Jovanovic_23,Jovanovic_24a,Jovanovic_24b}.
The GRAVITY collaboration  used  observational data covering the period from 1992 to 2022 to obtain  further constraints on Yukawa gravity  theory
(or in other words, constraints on fifth force parameters) \cite{GRAVITY_25}.
A brief review of different ways to constrain graviton mass was presented in \cite{Zakharov_18c}.

Using cosmological observational data that the Universe should be very close to flat and cosmological solutions obtained within the framework of the relativistic theory of gravity,  it was found in 
\cite{Gershtein_04,Logunov_06}
 \begin{gather}
 \Omega_{tot} = 1+ f^2/6,
 \label{RTG_Cosmology} 
\end{gather}
{where $\Omega_{tot}$ is a dimensionless total mass density in critical density units},       $f=m_g/m^0_H$,  $m^0_H=\dfrac{\hbar H_0}{c^2}=3.8 h_{100} \times 10^{-66}~{\rm g} \approx 1.3 $ $h_{100}   \times 10^{-33}~{\rm eV}/{\rm c^2}$
  is a Hubble mass  \cite{Gershtein_04}, while $H_0=h_{100} \times 100  ({\rm km/s})/{\rm Mpc}$ is a current Hubble constant ($h_{100}$ is a useful dimensionless constant).
{Using the upper observational constraint on $\Omega_{tot}$ the authors concluded that the graviton mass should be less than $m_g < 1.3 \times 10^{-66}~{\rm g} \approx 9 \times 10^{-34}~{\rm eV}/{\rm c}^2$ (for $h_{100} \approx 0.7$) and the Universe must be closed.
This graviton mass constraint is one of the strongest among bounds given in PDG}.

{
One of the strongest non-cosmological constraint for the graviton mass $m_g < 6.76 \times 10^{-23}$~eV$/{\rm c^2}$ (90\% C. L.) was found from planetary ephemerides  INPOP17a
\footnote{\url{https://www.imcce.fr/content/medias/recherche/equipes/asd/inpop/inpop17a_v3_090817.pdf}}    \cite{bern19}. 
Soon after that the authors used  new data on ephemerides INPOP19a \footnote{\url{https://www.imcce.fr/content/medias/recherche/equipes/asd/inpop/inpop19a_20191214.pdf}} and added observational data for Mars orbiters,
Cassini, Messenger, and Juno and the graviton mass constraint was significantly improved
$m_g < 3.62  \times 10^{-23}$~eV$/{\rm c^2}$ (99.7\% C. L.) \cite{Bernus_20}.
}

\section{Results}

\subsection{Constraining graviton mass from precession of S-stars around the Galactic Center}

In order to obtain new constraints on graviton mass using the observed orbits of S-stars around the Galactic Center (GC), we use a phenomenological modification of the Newtonian gravitational potential with a non-linear Yukawa-like correction  \cite{Will_98,will18}. Various experimental constraints on the graviton mass may be found in \cite{zyla20}.

There exists a wide range of massive gravity theories \cite{deRham_17}, and some models among them predict that gravitational potential in the Newtonian limit acquires a Yukawa suppression \cite{deRham_17}. In that case the Poisson equation for Newtonian gravity $\nabla ^{2}\Phi =4\pi G\rho$ is modified by graviton mass $m_g$, and it then takes the following form \cite{pois14}:
\begin{equation}
\left(\nabla^2+\dfrac{1}{\lambda^2}\right)\Phi=4\pi\,G\rho,
\label{eq:poisson}
\end{equation}
where we have $\lambda$:
\begin{equation}
\lambda = \dfrac{h}{m_g\,c},
\label{eq:compton}
\end{equation}
which is the Compton wavelength of the graviton \cite{Will_98,will18,pois14}). In such a case the spherically symmetric
potential $\Phi$ of a body of mass $M$ is given by:
\begin{equation}
\Phi\left(r\right)=-\dfrac{GM}{r}\, e^{-\dfrac{r}{\lambda}}.
\label{eq:potential}
\end{equation}

We also assume that the metric at leading order in the Newtonian regime would be  \cite{bern19}:
\begin{equation}
ds^2 = \left( -1 + \dfrac{2GM}{c^2 r}~e^{-\dfrac{r}{\lambda}} \right) c^2 dt^2 + \left( 1 + \dfrac{2GM}{c^2 r}~e^{-\dfrac{r}{\lambda}} \right) dl^2,
\quad dl^2 \equiv dx^2+dy^2+dz^2.
\label{eq:metric}
\end{equation}

In order to derive the expression for orbital precession in the gravitational potential (\ref{eq:potential}), we make an assumption that it does not differ significantly from the Newtonian potential $\Phi_N(r)=-\dfrac{GM}{r}$. Orbital precession $\Delta\varphi$ per orbital period, induced by small perturbations to the Newtonian gravitational potential which are described by the perturbing potential $V(r)=\Phi(r)-\Phi_N(r)$, could be evaluated as \cite{Zakharov_18}:
\begin{equation}
\Delta\varphi^{rad} = \dfrac{-2L}{GM e^2}\int\limits_{-1}^1
{\dfrac{z\cdot dz}{\sqrt{1 - z^2}}\dfrac{dV\left( z \right)}{dz}},
\label{eq:precorb}
\end{equation}
\noindent where we have $r$ related to $z$ via $r = \dfrac{L}{1 + ez}$, and we have $L = a\left( {1 - {e^2}} \right)$ {the focal parameter} of the orbital ellipse
{($a$ is semi-major axis, and $e$ is eccentricity of the orbit)}.

\subsubsection{Stellar orbits in extended/modified PPN formalisms}

{We used the parameterized post-Newtonian (PPN)\index{parameterized post-Newtonian (PPN)}\footnote{{See, book \cite{Will_93} for details on PPN-approach.}} equations of motion with aim to simulate the orbits of S-stars around GC. The PPN formalism is very important tool for assessing the viability of gravity theories. Using this tool one can compare the predictions of a given theory of gravity to observations into two independent parts: the derivation of a number of parameter values from any given theory of gravity, and the constraint of these values by astrophysical observations in week field limit.}

{For situations in which gravity is very weak, GR and the Newton theory lead to very similar predictions for the motion of bodies and the propagation of light. One can start with the Newtonian description and then, adding correction
terms (first the largest effect, and then smaller one) that take into account the effects of GR.
The PN formalism is a method for performing those corrections. The largest effects are called of  first post-Newtonian order, 1PN, the next smallest ones of second order, and so on. The progression
of ever smaller corrections is also called the PN expansion. Higher-order terms can be added to increase accuracy, but for strong fields. The PN formalism is a calculation tool that expresses Einstein's (nonlinear) equations of gravity in terms of the lowest-order deviations from Newton's law of universal gravitation.
The parameterized post-Newtonian formalism (PPN formalism), is a version of this formulation that explicitly includes the parameters in which a general theory of gravity can differ from Newtonian gravity. It is used as a tool to compare the Newtonian and Einstein gravity in the limit in which the gravitational field is weak and generated by objects moving slowly compared to the speed of light. The PPN formalism can be applied to most metric theories of gravitation with only a few exceptions.}

{PPN formalism completely characterizes the weak-field behavior of a gravity theory by a set of 10 PPN parameters in which this theory differs from Newtonian gravity. These parameters are:
\begin{itemize}
\item $\gamma$ - measuring how much of the space-curvature is produced by unit rest mass,
\item $\beta$ - measuring how much of "nonlinearity" is in the superposition law for gravity,
\item $\xi$ - measuring preferred-location effects,
\item $\alpha_1,\alpha_2,\alpha_3$ - measuring preferred-frame effects,
\item $\zeta_1, \zeta_2, \zeta_3, \zeta_4$ - measuring violation of conservation of total momentum.
\end{itemize}
In GR $\gamma$ and $\beta$ are equal to 1, while remaining PPN parameters are equal to 0. In alternative theories of gravity at least some of a PPN parameters should differ in respect to the GR.}

{Metric tensor $g_{\mu\nu}$ can be then represented as expansion over 10 dimensionless metric potentials ($U$, $U_{ij}$, $\Phi_{W}$, $A$, $\Phi_{1}$, $\Phi_{2}$, $\Phi_{3}$, $\Phi_{4}$, $V_{i}$, $W_{i}$) \cite{Will_14} as:}
\begin{align*}
g_{00} = & - 1 + 2 U - 2 \beta U^2 - 2 \xi \Phi_W +\\
& (2 \gamma + 2 + \alpha_3 + \zeta_1 - 2 \xi ) \Phi_1 + \ldots + {\cal O} (\epsilon^3)\\
g_{0i} = & - \frac{1}{2} (4 \gamma + 3 + \alpha_1 - \alpha_2 + \zeta_1 - 2 \xi ) V_i -\\
& \frac{1}{2} (1 + \alpha_2 - \zeta_1 + 2 \xi) W_i - \ldots + {\cal O} (\epsilon^{5/2})\\ 
g_{ij} = & (1 + 2 \gamma U) \delta_{ij} + {\cal O} (\epsilon^2).
\end{align*}

{Using the PPN formalism a certain gravitation phenomena can be expressed in a very simple way. For example, PPN expression for precession angle per orbit is $\Delta\phi\approx \left(2 + 2 \gamma - \beta\right){\frac {2\pi GM}{c^{2}a\left(1-e^{2}\right)}}$, while for light deflection angle it is $\alpha\approx 2\left(1+\gamma\right)\frac{GM}{c^2\xi}$.}

{However, the massive gravity theories with Yukawa-like potentials belong to rare exceptions which cannot be described by the above standard PPN formalism because Newtonian order terms are modified by the presence of massive fields, so that the Newtonian potential acquires a Yukawa-like correction. Therefore, we used so-called extended PPN formalism which includes both Yukawa and the post-Newtonian corrections \cite{Jovanovic_23}.}

As it is well known that Yukawa-like potentials could not be entirely represented by the standard PPN formalism, these potentials require its extension/modification (see the related references in \cite{Jovanovic_23}). This is also valid for the potential (\ref{eq:potential}), as well as its corresponding metric (\ref{eq:metric}). Since Yukawa gravity is indistinguishable from GR up to the first post-Newtonian correction, in addition to the standard PPN equations of motion $\vec{\ddot{r}}_{\scriptscriptstyle GR}$ in GR, PPN equations of motion $\vec{\ddot{r}}_{\scriptscriptstyle Y}$ in potential (\ref{eq:potential}) also include an additional term $\vec{\ddot{r}}_{\scriptscriptstyle\lambda}$ with exponential correction due to the perturbing potential $V(r)$. In such extended PPN formalism (denoted here as $\mathrm{PPN_Y}$), the equations of motion are:

\begin{equation}
\vec{\ddot{r}}_{\scriptscriptstyle Y} = \vec{\ddot{r}}_{\scriptscriptstyle GR} + \vec{\ddot{r}}_{\scriptscriptstyle\lambda},
\qquad \vec{\ddot{r}}_{\scriptscriptstyle GR} = \vec{\ddot{r}}_{\scriptscriptstyle N} + \vec{\ddot{r}}_{\scriptscriptstyle PPN},
\label{eq:eom0}
\end{equation}

\noindent where the $\vec{\ddot{r}}_{\scriptscriptstyle N}$ is the Newtonian acceleration, $\vec{\ddot{r}}_{\scriptscriptstyle PPN}$ is the first post-Newtonian correction and $\vec{\ddot{r}}_{\scriptscriptstyle\lambda}$ is additional Yukawa correction. They are given by the following expressions, respectively:
\begin{equation}
\begin{array}{l}
\vec{\ddot{r}}_{\scriptscriptstyle N} = -GM \dfrac{\vec{r}}{r^3}, \\
\\
\vec{\ddot{r}}_{\scriptscriptstyle PPN} = \dfrac{GM}{c^2r^3} \left[\left(4\dfrac{G M}{r}-\vec{\dot{r}}\cdot\vec{\dot{r}}\right) \vec{r} + 4\left(\vec{r}\cdot\vec{\dot{r}}\right)\vec{\dot{r}}\right], \\
\\
\vec{\ddot{r}}_{\scriptscriptstyle\lambda} = GM \left[ 1 - \left(1 + \dfrac{r}{\lambda} \right) e^{-\dfrac{r}{\lambda}} \right] \dfrac{\vec{r}}{r^3}.
\end{array}
\label{eq:ppn}
\end{equation}
The additional Yukawa correction $\vec{\ddot{r}}_{\scriptscriptstyle\lambda}$ becomes negligible when $\lambda\rightarrow\infty$,
and then $\vec{\ddot{r}}_{\scriptscriptstyle Y}\rightarrow\vec{\ddot{r}}_{\scriptscriptstyle GR}$, i.e. PPN equations
of motion $\vec{\ddot{r}}_{\scriptscriptstyle Y}$ in potential (\ref{eq:potential}) reduce to the standard PPN equations of motion in GR.

The orbital precession in GR\index{orbital precession in GR} is given by the well known expression for the Schwarzschild precession \cite{Will_14}:
\begin{equation}
\Delta\varphi_{\scriptscriptstyle GR}^{rad}\approx\dfrac{6\pi G M}{c^2 a(1-e^2)}.
\label{eq:precgr}
\end{equation}
Recently, the GRAVITY Collaboration detected the orbital
precession of the S2 star around the SMBH at GC and showed that it was close to the above prediction of GR \cite{GRAVITY_20}.
For that purpose, they introduced $f_{SP}$, an ad hoc factor, in front of the first PN correction of GR in order to parameterize the effect of the Schwarzschild metric. In such modified PPN formalism (denoted here as $\mathrm{PPN_{SP}}$),
the equations of motion are given by:
\begin{equation}
\vec{\ddot{r}}_{\scriptscriptstyle SP} = \vec{\ddot{r}}_{\scriptscriptstyle N} + f_{\scriptscriptstyle SP}\cdot\vec{\ddot{r}}_{\scriptscriptstyle PPN}.
\label{eq:eomfsp}
\end{equation}
Then, the corresponding modified expression for the Schwarzschild precession is \cite{GRAVITY_20}:
\begin{equation}
\Delta\varphi_{\scriptscriptstyle SP}^{rad}=f_{\scriptscriptstyle SP}\cdot\Delta\varphi_{\scriptscriptstyle GR}^{rad}.
\label{eq:precfsp}
\end{equation}

For $f_{SP}=1$, the expression (\ref{eq:eomfsp}) reduces to the standard PPN equations of motion $\vec{\ddot{r}}_{\scriptscriptstyle GR}$ in GR. The parameter $f_{SP}$ shows to which extent some gravitational model is relativistic, and it is defined with: $f_{SP} = (2 + 2\gamma - \beta) / 3$, where $\beta$ and $\gamma$ are the post-Newtonian parameters, and in the case of GR the both of them equals to 1 (and thus $f_{SP}=1$ in this case). For $f_{SP}=0$ we have the Newtonian case recovered. However, in the case of S2 star the value of $f_{SP} = 1.10 \pm 0.19$ was obtained by the GRAVITY Collaboration \cite{GRAVITY_20}. Recently, this collaboration updated the above first estimate to $f_{SP} = 0.85 \pm 0.16$ \cite{abut22}. Besides, they also presented the following estimate obtained from the fit by using data of four stars (S2, S29, S38, S55), with the $1\sigma$ uncertainty: $f_{SP} = 0.997 \pm 0.144$ \cite{abut22}.

Here we want to constrain the parameter $\lambda$ in that way the orbital precession in the gravitational potential (\ref{eq:potential}) deviates from the Schwarzschild precession in GR by a factor $f_{SP}$. It can be achieved if we provide that the total orbital precession in Yukawa gravity\index{orbital precession in Yukawa gravity}, given by the sum $\Delta\varphi_{\scriptscriptstyle GR}+\Delta\varphi_{\scriptscriptstyle Y}$, is close to the observed precession $\Delta\varphi_{\scriptscriptstyle SP}$ (which is obtained by the GRAVITY Collaboration):
\begin{equation}
\Delta\varphi_{\scriptscriptstyle Y} + \Delta\varphi_{\scriptscriptstyle GR} \approx \Delta\varphi_{\scriptscriptstyle SP}
\quad\Leftrightarrow\quad
\pi\sqrt{1-e^2}\dfrac{a^2}{\lambda^2} + \dfrac{6\pi GM}{c^2 a(1 - e^2)} \approx f_{SP}\dfrac{6\pi GM}{c^2 a(1 - e^2)}.
\label{eq:prectot}
\end{equation}
If we take into account the third Kepler law (since the orbits are almost Keplerian):
\begin{equation}
\dfrac{P^2}{a^3} \approx \dfrac{4\pi^2}{GM},
\label{eq:kepler}
\end{equation}
and using Eqs. (\ref{eq:prectot}) and (\ref{eq:kepler}), then we can obtain the following relation between $\lambda$ and $f_{SP}$:
\begin{equation}
\lambda(P,e,f_{SP}) \approx \dfrac{cP}{2\pi}\dfrac{(1-e^2)^{\frac{3}{4}}}{\sqrt{6(f_{SP} - 1)}},\qquad f_{SP} > 1.
\label{eq:lambda}
\end{equation}
As $\Delta\varphi_{\scriptscriptstyle Y}$ always gives a
positive contribution to the total precession in Eq. (\ref{eq:prectot}), this derived condition can be used for constraining the Compton wavelength $\lambda$ of the graviton by the observed values of $f_{SP}$ only in the cases when $f_{SP}$ is larger than 1. Then, the corresponding constraints on the graviton mass $m_g$ can be found by inserting the obtained value of $\lambda$ into Eq. (\ref{eq:compton}). The relative error of the parameter $\lambda$, as well as of the graviton mass $m_g$, can be found in this case by differentiating the logarithm of the above expression (\ref{eq:lambda}):
\begin{equation}
\dfrac{|\Delta\lambda|}{\lambda}=\dfrac{|\Delta m_g|}{m_g} \leq \left(\dfrac{\left|\Delta P\right|}{P}+\dfrac{3 e\left|\Delta e\right|}{2 (1-e^2)}+\dfrac{\left|\Delta f_{SP}\right|}{2 (f_{SP}-1)}\right).
\label{eq:relerr}
\end{equation}
It can be seen that the potential (\ref{eq:potential}) results with the same relative errors as the corresponding Yukawa potential derived in the frame of $f(R)$
theories of gravity (see e.g. Ref. \cite{Jovanovic_24b}).

In our investigation we have two main assumptions: (1) the parameter $\Lambda$ of the Yukawa-like gravitational potential (representing the range of interaction) corresponds to the graviton Compton wavelength $\lambda_g$, and (2) the orbital precession in Yukawa gravity should be close to the prediction of GR for Schwarzschild precession. These assumptions enabled us to obtain the constraints on the graviton mass $m_g$ in two different ways. First way is by use of the factor $f_{SP}$ under which the observed orbital precession deviates from the Schwarzschild precession in GR, and second way is by direct fitting of the simulated orbits in Yukawa gravity into the corresponding observed orbits of S-stars. In the first approach, we derived the relation between the parameter $\Lambda$ and $f_{SP}$ when universal constant is $\delta=1$ (see equation (\ref{eq:lambda})), with precession which deviates from GR according to $f_{SP}$ \cite{Jovanovic_23,Jovanovic_24a,Jovanovic_24b}. After taking that the graviton Compton wavelength $\lambda_g=\Lambda$, we obtained the corresponding estimates for the graviton mass $m_g$ in the case of investigated S-stars \cite{Jovanovic_24a}. The value of upper bound for graviton mass $m_g$, in the case of S2 star which is obtained for the best-fit value of $f_{SP}$ given in GRAVITY Collaboration (2020) \cite{GRAVITY_20} $f_{SP} = 1.10 \pm 0.19$, is $m_g < (124.9\pm 120.2)\times 10^{-24}~$eV$/{\rm c^2}$. The value of upper bound for graviton mass $m_g$ in the case of S38 star (we also take the same $f_{SP}$) is $m_g < (76.9\pm 73.3)\times 10^{-24}~$eV$/{\rm c^2}$ \cite{Jovanovic_24a}.

\subsubsection{PPN fit of the observed orbits of S2 and S38 stars}

The second approach to obtain the constraints on the graviton mass $m_g$ is direct orbital fitting in the frame of Yukawa gravity. We estimated the value of the Compton wavelength $\lambda$ of the graviton by fitting the simulated orbits in the extended $\mathrm{PPN_Y}$ formalism into the observed orbits of S2 and S38 stars. In order to calculate this, we used the publicly available astrometric observations of S2 and S38 stars from \cite{gill17}.

\begin{table}[ht!]
\centering
\caption{Best-fit values of the graviton Compton wavelength $\lambda$, SMBH mass $M$, distance $R$ to the GC and the osculating orbital elements $a, e, i, \Omega, \omega, P, T$ of the S2 star orbit.}
\label{tab1}
\begin{tabular}{cllll}
\hline
\\
~~~~~Parameter~~~~~~~~~~& Value~~~~~~~~~~~~~~~ & Fit error~~~~~~~~~~~~~~~ & Unit~~~~~~~~~~ \\
\\
\hline
\\
$\lambda_g$& 58000 & 9800 & AU \\
$M$      & 4.10 & 0.579 & $10^6\,M_\odot$ \\
$R$      & 8.30 & 0.246 & kpc \\
$a$      & 0.1229 & 0.00527 & arcsec \\
$e$      & 0.8787 & 0.01213 & \\
$i$      & 134.90 & 2.049 & $^\circ$ \\
$\Omega$ & 224.51 & 6.840 & $^\circ$ \\
$\omega$ & 62.70 & 5.781 & $^\circ$ \\
$P$      & 16.05 & 0.541 & yr \\
$T$      & 2018.29626 & 1.629346 & yr \\
\\
\hline
\noalign{\smallskip} \noalign{\smallskip}
\end{tabular}
\end{table}

\begin{figure}[ht!]
\centering
\includegraphics[width=0.90\textwidth]{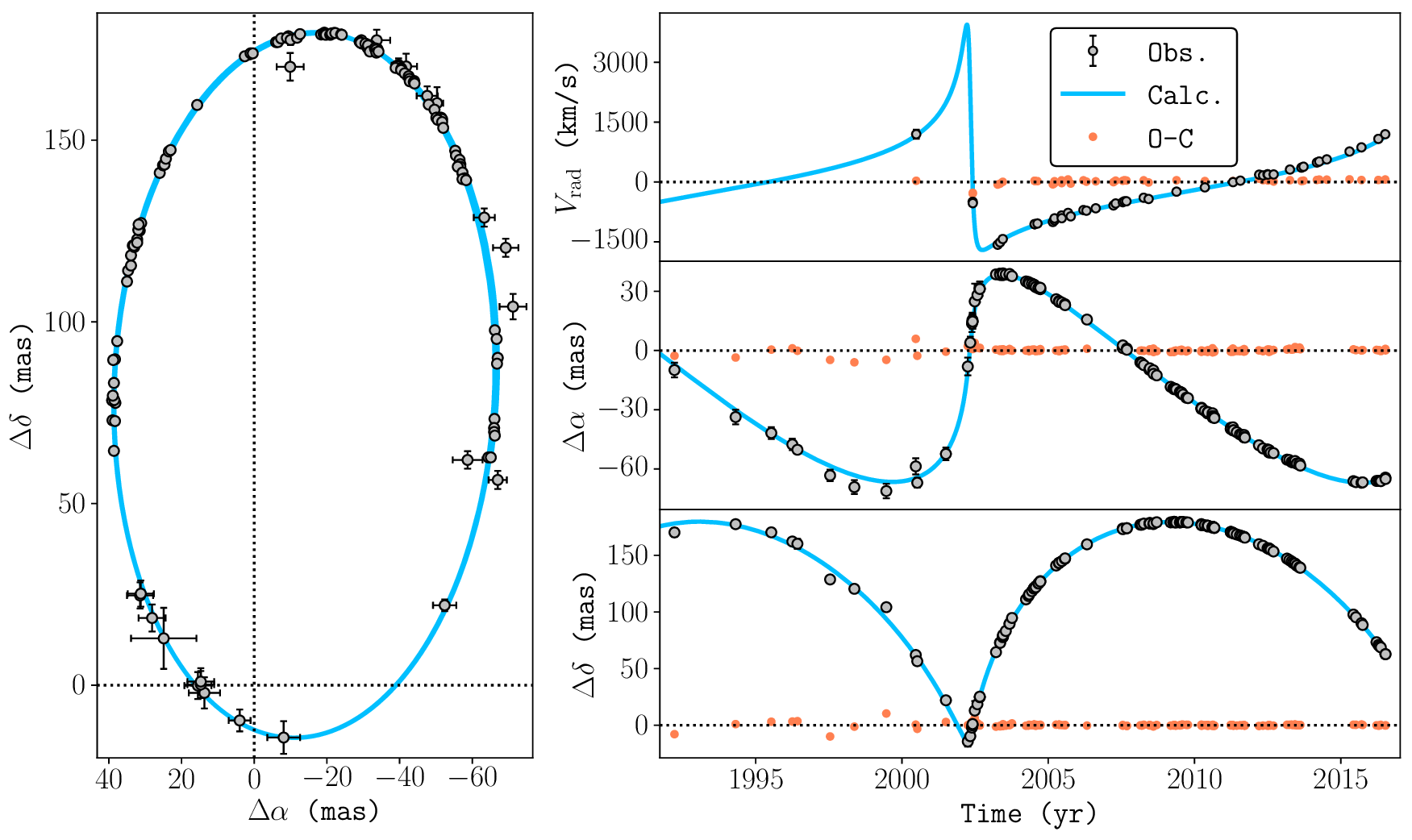}
\caption{(Color online) \textit{Left:} Comparison between the best-fit orbit of the S2 star (blue solid line), simulated in the extended
$\mathrm{PPN_Y}$ formalism, and the corresponding astrometric observations from \cite{gill17} (black circles with error bars).
\textit{Right:} The same for the radial velocity of the S2 star (top), as well as for its $\alpha$ (middle) and $\delta$
(bottom) offset relative to the position of Sgr A* at the coordinate origin. Red dots in the right panels denote the
corresponding O-C residuals}.
\label{fig1}
\end{figure}

\begin{table}[ht!]
\centering
\caption{Best-fit values of the graviton Compton wavelength $\lambda$, SMBH mass $M$, distance $R$ to the GC and the osculating orbital elements $a, e, i, \Omega, \omega, P, T$ of the S38 star orbit.}
\label{tab2}
\begin{tabular}{cllll}
\hline
\\
~~~~~Parameter~~~~~~~~~~& Value~~~~~~~~~~~~~~~ & Fit error~~~~~~~~~~~~~~~ & Unit~~~~~~~~~~ \\
\\
\hline
\\
$\lambda_g$& 79439.2 & 40583.07  & AU \\
$M$      & 4.30 & 0.28 & $10^6\,M_\odot$ \\
$R$      & 8.18 & 0.020 & kpc \\
$a$      & 0.1407 & 0.00125 & arcsec \\
$e$      & 0.8230 & 0.01048 & \\
$i$      & 177.75 & 6.294 & $^\circ$ \\
$\Omega$ & 101.80 & 0.997 & $^\circ$ \\
$\omega$ & 17.86 & 0.910 & $^\circ$ \\
$P$      & 19.17 & 0.014 & yr \\
$T$      & 2003.49562 & 0.027371 & yr \\
\\
\hline
\noalign{\smallskip} \noalign{\smallskip}
\end{tabular}
\end{table}

\begin{figure}[ht!]
\centering
\includegraphics[width= 0.90\textwidth]{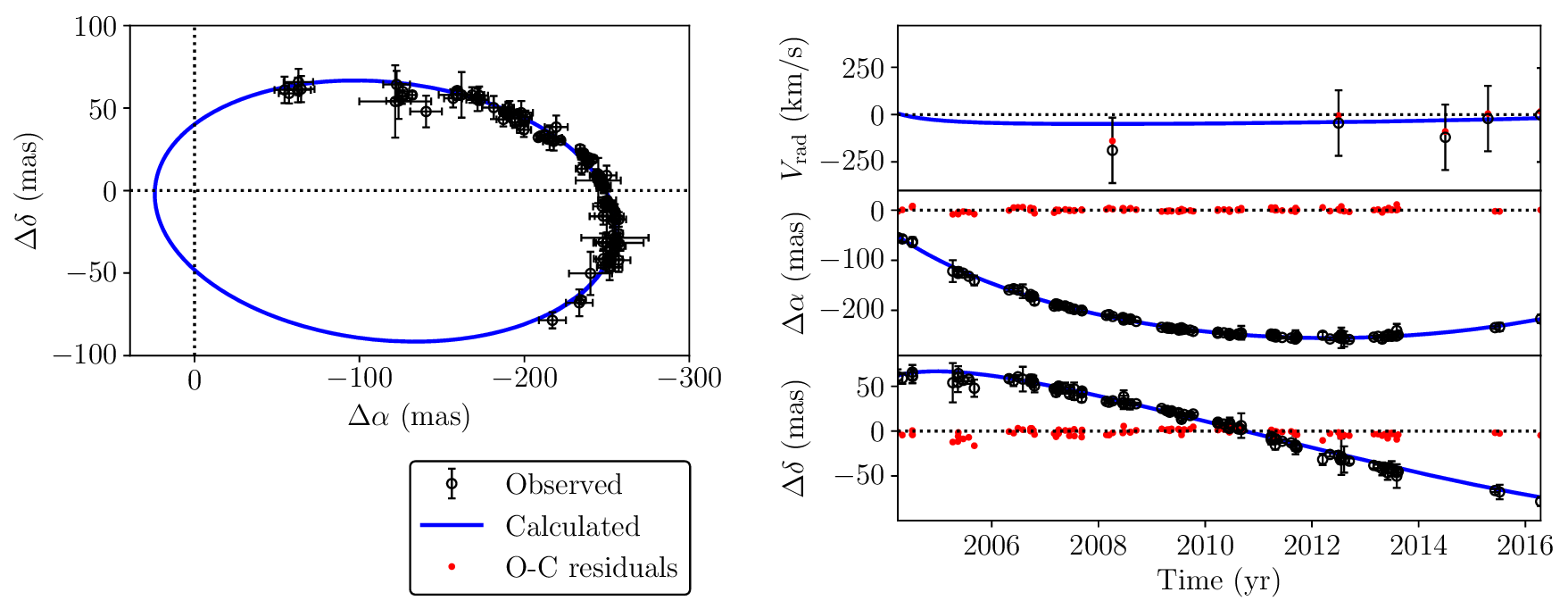}
\caption{(Color online) \textit{Left:} Comparison between the best-fit orbit of the S38 star (blue solid line), simulated in the extended
$\mathrm{PPN_Y}$ formalism, and the corresponding astrometric observations from \cite{gill17} (black circles with error bars).
\textit{Right:} The same for the radial velocity of the S38 star (top), as well as for its $\alpha$ (middle) and $\delta$
(bottom) offset relative to the position of Sgr A* at the coordinate origin. Red dots in the right panels denote the
corresponding O-C residuals.}
\label{fig2}
\end{figure}

Orbital fitting in the frame of extended $\mathrm{PPN_Y}$ formalism we performed by minimization of the reduced $\chi^2$ statistics:
\begin{equation}
\chi_{\mathrm{red}}^2 = \dfrac{1}{2\left(N-\nu\right)}{\sum\limits_{i = 1}^N {\left[ {{{\left( {\dfrac{x_i^o - x_i^c}{\sigma_{xi}}}
\right)}^2} + {{\left( \dfrac{y_i^o - y_i^c}{\sigma_{yi}} \right)}^2}} \right]} },
\label{eq:chi2}
\end{equation}
where $(x_i^o, y_i^o)$ is the $i$-th observed position and $(x_i^c, y_i^c)$ is the estimated position; $N$
is the number of observations, $\nu$ is a number of unknown parameters, while $\sigma_{xi}$ and $\sigma_{yi}$ are the observed astrometric uncertainties.

We have found the values of the graviton Compton wavelength $\lambda$, SMBH mass $M$, distance $R$ to the GC and the osculating orbital elements $a, e, i, \Omega, \omega, P, T$ which correspond to the minimum of $\chi_{\mathrm{red}}^2$. We managed this using the differential evolution optimization method, implemented as Python Scipy function:
\href{https://docs.scipy.org/doc/scipy/reference/generated/scipy.optimize.differential_evolution.html}{\nolinkurl{scipy.optimize.differential\_evolution}} .

The obtained results of the orbital fitting in the case of S2 star are presented in Fig. \ref{fig1}, and their corresponding best-fit values of the parameters are given in Table \ref{tab1}. As it can be clearly seen from Fig. \ref{fig1}, the best-fit orbit of S2 star in the extended $\mathrm{PPN_Y}$ formalism is in a very good agreement with the observations. Regarding the graviton Compton wavelength $\lambda$ obtained from S2 star orbit, it can be concluded that its best-fit value is $\lambda_g \approx 5.8\times 10^4$~AU. The obtained results of the orbital fitting in the case of S38 star are presented in Fig. \ref{fig2} and their corresponding best-fit values of the parameters are given in Table \ref{tab2}. By inspection of Fig. \ref{fig2}, like in a previous case, it can be noticed that the best-fit orbit of S38 star in the extended $\mathrm{PPN_Y}$ formalism is in a very good agreement with the observations. The best-fit value of graviton Compton wavelength $\lambda$ obtained from S38 star orbit is $\lambda_g \approx 7.9\times 10^4$~AU. 
In the case of S2 star we obtain following value for upper bound of graviton mass $m_g = (142.9 \pm 24.1) \times 10^{-24}~{\rm eV}/{\rm c}^2$, and in the case of S38 star it is $m_g < (92\pm 47)\times 10^{-24}~{\rm eV}/{\rm c}^2$.

We can see that our constraints on the graviton mass obtained using the above two methods are independent, but consistent with the corresponding LIGO's result of $m_g\le 1.2\times 10^{-22}~{\rm eV}/{\rm c}^2$ estimated from the first gravitational wave signal GW150914 \cite{LIGO_16}.

\section{Conclusion} 
In order to obtain the constraints on the graviton mass $m_g$ we used the phenomenological Yukawa-like gravitational potential from \cite{Will_98,will18}. The results from this part of our research can be summarized as follows:

\begin{enumerate}
\item We used two different approaches based on: (i) recovering the recently detected Schwarzschild precession in the orbit of S2 star using the Yukawa-like gravitational potential and (ii) direct orbital fitting;
\item We used phenomenological Yukawa-like gravitational potential (\ref{eq:potential}), and found the condition for its parameter $\lambda$ under which the orbital precession in this potential deviates from the Schwarzschild precession in GR by a factor $f_{SP}$;
\item The relation (\ref{eq:lambda}) derived from the phenomenological potential (\ref{eq:potential}) in the frame of the two modified/extended PPN formalisms could be very useful to obtain the reliable constraints on the graviton mass $m_g$ from the latest estimates for $f_{SP}$ by the GRAVITY Collaboration, in the cases when $f_{SP} > 1$;
\item The both approaches resulted with the constraints on the graviton mass which are independent, but consistent with the corresponding LIGO’s estimate;
\item These results were also confirmed in the case of S2 and S38 stars in the frame of the extended
$\mathrm{PPN_Y}$ formalism. We obtained the best-fit value for the graviton Compton wavelength $\lambda$ within
the error intervals of its corresponding estimates calculated according to Eq. (\ref{eq:lambda}) from the detected values of $f_{SP}$;
\item In case of using Yukawa-like gravitational potential (for $f_{SP} = 1.10 \pm 0.19$ estimated by the GRAVITY Collaboration \cite{GRAVITY_20}), we obtained the value of upper bound for graviton mass $m_g$. In the case of S2 star it is $m_g < (124.9\pm 120.2)\times 10^{-24}~$eV$/{\rm c^2}$. The value of upper bound for graviton mass $m_g$ in the case of S38 star (we also take the same $f_{SP}$) is $m_g < (76.9\pm 73.3)\times 10^{-24}~$eV$/{\rm c^2}$ \cite{Jovanovic_24a};
\item In case of using direct orbital fitting, we obtained that the upper bounds on graviton mass $m_g$ slightly larger, but still within the error intervals of the corresponding values from the first approach. In the case of S2 star it is $m_g = (142.9 \pm 24.1) \times 10^{-24}~$eV$/{\rm c^2}$. In the case of S38 star it is $m_g < (92\pm 47)\times 10^{-24}~$eV$/{\rm c^2}$. In any case, more precise future observations would be useful in order to further improve the upper graviton mass bounds.
\end{enumerate}

We believe that the obtained results also demonstrate that the existence of massive graviton could play a fundamental role in modern physics, and that it is important to further investigate this topic.

\section*{Acknowledgments}

{
AFZ appreciates the Organizing Committee of the VIII International Conference "Models in Quantum Field Theory" for an invitation to present an invited talk on the subject at the Conference.}
{PJ, DB and VBJ acknowledge the support of Ministry of Science, Technological Development and Innovations of the Republic of Serbia through the Project contracts No. 451-03-33/2026-03/200002 and 451-03-33/2026-03/200017.
}

\section*{Conflicts of interest} 
The authors declare no conflicts of interest.

\end{document}